\documentclass[paper]{JHEP}
\usepackage{epsfig}
\def\as{\alpha_{\mbox{\tiny S}}}
\def\alb{\bar\as}

\def\gE{\gamma_{\mbox{\tiny E}}}
\def\DL{{\mbox{\scriptsize DL}}}

\def\mR{\mu_{\mbox{\tiny R}}}
\def\res{\,{\mbox{\scriptsize jet}}}
\def\om{\omega}
\def\alom{\frac{\alb}{\om}}
\def\omal{\frac{\om}{\alb}}
\def\glip{\gamma_{\mbox{\tiny L}}}
\def\VEV#1{\overline{#1}}
\def\sig{\sigma}
\title{\boldmath Multiplicity of (Mini-)Jets at Small $x$
\footnote{Work supported in part by the UK Particle Physics and
Astronomy Research Council and by the EU Fourth Framework Programme
`Training and Mobility of Researchers', Network `Quantum Chromodynamics
and the Deep Structure of Elementary Particles',
contract FMRX-CT98-0194 (DG 12 - MIHT).}}
\author{Carlo Ewerz$^{a,b}$ and Bryan R.\ Webber$^{c,a}$\\
$^a$Cavendish Laboratory, University of Cambridge,\\
Madingley Road, Cambridge CB3 0HE, U.K.\\[1mm]
$^b$DAMTP, University of Cambridge,\\
Silver Street, Cambridge CB3 9EW, U.K.\\[1mm]
$^c$Theory Division, CERN,\\
CH-1211 Geneva 23, Switzerland\\[1mm]
E-mail: \email{carlo@hep.phy.cam.ac.uk}, \email{webber@hep.phy.cam.ac.uk}}
\abstract{We derive closed expressions for the mean and variance of the
(mini-)jet multiplicity distribution in hard scattering processes at low $x$.
Here (mini-)jets are defined as those due to initial-state radiation of
gluons with transverse momenta greater than some resolution scale $\mR$,
where $\Lambda^2\ll\mR^2\ll Q^2$, $\Lambda$ being the intrinsic QCD scale
and $Q$ the momentum transfer scale of the hard scattering.
Our results are valid to leading order in $\log(1/x)$ but
include all sub-leading logarithms of $Q^2/\mR^2$.
As an illustration, we predict the mini-jet multiplicity in
Higgs boson production at the Large Hadron Collider.}
\preprint{Cavendish--HEP--99/05\\CERN--TH/99--220\\DAMTP--1999--79\\

hep-ph/9907430
}
\begin{document}
\section{Introduction}
The prediction of jet multiplicities in hard scattering processes
represents an important, well-defined challenge to perturbative QCD.
Especially challenging is the calculation of the jet multiplicity at
small $x$, that is, in processes where the hard-scattering scale, $Q^2$,
is much less than the overall collision energy-squared $s$
($x\sim Q^2/s\ll 1$).  Such processes include
not only the familiar case of deep inelastic lepton scattering
at low values of the Bjorken variable, but also Higgs and gauge
boson production at high-energy hadron colliders
(see e.g.\ ref.~\cite{ESW}).
A large associated jet multiplicity could pose
serious triggering and background problems for experiments
searching for new physics at the future Large Hadron Collider (LHC).

The particular problem at small $x$ is that higher-order perturbative
contributions may become enhanced by factors of $\ln(1/x)$. 
As a first approximation, one can consider only those terms
that are enhanced by a factor of $\ln(1/x)$
for each power of the strong coupling $\as$. In this approximation,
the fully inclusive deep inelastic structure functions satisfy the
leading-order Balitskii-Fadin-Kuraev-Lipatov (BFKL) equation \cite{BFKL,FR}.

When computing the associated multi-jet production rates at small
$x$ \cite{FSV,Web98,EW1}, one encounters another potentially
large logarithmic factor, namely  $\ln(Q^2/\mR^2)$ where
$\mR$ is the minimum transverse momentum that a jet must have
in order to be resolved.  For mini-jets, by which we
mean jets with transverse momenta that are relatively low
compared with the hard scattering scale, it clearly becomes
important to resum terms enhanced by factors of $\ln(Q^2/\mR^2)$
as well as $\ln(1/x)$.

In ref.~\cite{Web98}, such a resummation was performed in the
double-logarithmic (DL) approximation, i.e.\ keeping only terms
of the form $[\as\ln(1/x)\,\ln(Q^2/\mR^2)]^n$ to all orders.
The result was remarkably simple, namely that the mean and
variance (mean square fluctuation) of the jet multiplicity
are respectively quadratic and cubic functions of $\ln(Q^2/\mR^2)$,
with simple coefficients.

In ref.~\cite{EW1} we extended the results of ref.~\cite{Web98}
concerning fixed-multiplicity jet rates to single logarithmic (SL)
accuracy, i.e.\ keeping all terms $[\as\ln(1/x)]^n\,[\ln(Q^2/\mR^2)]^m$
with $0<m\leq n$. We also showed that an iterative procedure could be used
to obtain the jet multiplicity moments to any desired order in $\as$, to
SL accuracy. However, we were not able to derive closed expressions
that resum the enhanced terms in the multiplicity moments to all orders.
That is done in the present paper.

An alternative approach to hard scattering at small $x$ is the 
CCFM equation \cite{C,CFM} based on angular ordering of gluon 
emissions. Remarkably, it can be shown \cite{Sal99} that 
for sufficiently inclusive observables the CCFM formalism 
leads to the same results as the BFKL equation. 
The jet multiplicities calculated in the present paper fall into 
this class of observables. The BFKL formalism turns out to be
technically simpler and therefore we adopt it here. 

The paper is organized as follows. 
First, in sect.~\ref{sec_genf}, we
solve the equation derived in ref.~\cite{EW1} for the jet-rate generating
function, by expressing the solution as an inverse Laplace transform.
In sect.~\ref{sec_mom} we obtain the jet multiplicity moments by
saddle-point evaluation of the inverse Laplace transform. The results
are discussed in sect.~\ref{sec_disc}.  We find that the simple
quadratic and cubic forms of the mean and variance found in the DL
approximation remain valid to SL precision. The coefficients, while
not particularly simple, can be expressed straightforwardly in terms
of the Lipatov anomalous dimension \cite{BFKL} and its derivatives.

As an illustrative application of our results, in sect.~\ref{sec_higgs}
we calculate the mean and variance of the mini-jet multiplicity in
Higgs boson production at the LHC, as functions of the Higgs mass.
 
\section{Jet-rate generating function}\label{sec_genf}
We write the contribution to the
gluon structure function at scale $Q^2$, $F(x,Q^2)$, in which
$r$ final-state mini-jets are resolved with transverse momenta
greater than $\mR$, in the form
\begin{equation}\label{Frx}
F^{(r\res)}(x,Q^2,\mR^2) = F(x,\mR^2)\otimes G^{(r)}(x,T)
\equiv\int_x^1\frac{dz}{z} F(z,\mR^2) G^{(r)}(x/z,T)\,,
\end{equation}
where the coefficient $G^{(r)}$ is a function of $x$ and
$T=\ln(Q^2/\mR^2)$ to be determined . The Mellin transformation
\begin{equation}\label{mellin}
f_\om(\ldots) = \int_0^1 dx\,x^\om f(x,\ldots)\;,
\end{equation}
with inverse
\begin{equation}\label{melinv}
f(x,\ldots) = \frac{1}{2\pi i}\int_C d\om\,x^{-\om-1}
f_\om(\ldots)\;,
\end{equation}
where the contour $C$ is parallel to the imaginary axis and
to the right of all singularities of the integrand,
converts the convolution in eq.~(\ref{Frx}) into a simple
product:
\begin{equation}\label{From}
F^{(r\res)}_\om(Q^2,\mR^2) = F_\om(\mR^2)\,G^{(r)}_\om(T)\;.
\end{equation}
Furthermore the evolution of the structure function in
$\om$-space is simply given by
\begin{equation}\label{Fevol}
F_\om(Q^2)=\exp[\glip(\alb/\om)T]F_\om(\mR^2) \,,
\end{equation}
where $\glip$ is the Lipatov anomalous dimension:
\begin{equation}\label{glip}
\glip(\alb/\om) = \alom+
2\zeta(3)\left(\alom\right)^4+
2\zeta(5)\left(\alom\right)^6+
12[\zeta(3)]^2\left(\alom\right)^7+\ldots\;,
\end{equation}
which is the solution of the equation
\begin{equation}\label{omlip}
\omega = -\alb\left[2\gE+\psi(\gamma)+\psi(1-\gamma)\right]
\equiv \alb\,\chi(\gamma)\;,
\end{equation}
where $\alb=3\as/\pi$, $\psi$ being the digamma function and
$\gE=-\psi(1)$ the Euler constant.

Introducing the jet-rate generating function
\begin{equation}\label{GuT}
G_\om(u,T)=\sum_{r=0}^\infty u^r G^{(r)}_\om(T)\;,
\end{equation}
we can write the $r$-jet rate, i.\,e.\ the fraction of events 
with $r$ mini-jets, as
\begin{equation}\label{Rrres}
R^{(r\res)}_\om(Q^2,\mR^2) = \frac{1}{r!}
\left.\frac{\partial^r}{\partial u^r}R_\om(u,T)\right|_{u=0}\;,
\end{equation}
where
\begin{equation}
R_\om(u,T)=\exp[-\glip(\alb/\om)T]\,G_\om(u,T)\;.
\end{equation}

We showed in ref.~\cite{EW1} that $G_\om(u,T)$ is given by
\begin{equation}\label{Gexpn}
G_\om(u,T)= 1+\int_0^T dt\,g_\om(u,t,T)\;,
\end{equation}
where the unintegrated function $g_\om(u,t,T)$
satisfies the integro-differential equation
\begin{equation}\label{gomu}
g_\om = \frac{u\alb}{\om+\alb t}\left[1+
\int_0^T dt'\,g_\om\left(u,\max\{t,t'\},T\right)
+2\sum_{m=1}^\infty\zeta(2m+1)\,
\frac{\partial^{2m}g_\om}{\partial t^{2m}}\right]\;.
\end{equation}
Rearranging terms and differentiating with respect to $t$, this gives
\begin{equation}\label{dgomu}
\frac{\partial}{\partial t}\left\{\left[1-(u-1)t\alom\right]g_\om\right\}
= -u\alom\left[1-2\sum_{m=1}^\infty\zeta(2m+1)\,
\frac{\partial^{2m+1}}{\partial t^{2m+1}}\right]g_\om\;.
\end{equation}
To find $g_\om$ we first recall the solution in double-logarithmic
approximation \cite{Web98},
\begin{equation}\label{gomDL}
g_\om^\DL(u,t,T)= u\alom\left[1-(u-1)t\alom\right]^{\frac{1}{u-1}}
\left[1-(u-1)T\alom\right]^{\frac{-u}{u-1}}\;.
\end{equation}
Notice that this expression vanishes, together with all its $t$-derivatives,
at the point $t=t_0=\om/[\alb(u-1)]$ as $u\to 1^+$. We shall assume
this is also true to SL accuracy, for some finite value of $t_0$,
and then verify that the solution is consistent with perturbation theory.
The equation (\ref{dgomu}) can now be solved by
introducing a Laplace transformation of the form
\begin{equation}\label{tildef}
\tilde g_\om(u,\gamma,T) = \int_{-\infty}^{t_0}
dt\,g_\om(u,t,T)\,e^{\gamma t}\;,
\end{equation}
with inverse
\begin{equation}\label{inverg}
g_\om(u,t,T) = \frac{1}{2\pi i}\int_\Gamma
d\gamma\,\tilde g_\om(u,\gamma,T)\,e^{-\gamma t}\;,
\end{equation}
where the contour $\Gamma$ is parallel to the imaginary axis and
to the right of all singularities of the integrand. This gives
\begin{equation}\label{tildeg}
\tilde g_\om(u,\gamma,T) = \tilde g_\om(u,\gamma_0,T)
\,e^{\phi_\om(u,\gamma)}\,,
\end{equation}
where
\begin{eqnarray}\label{phidef}
\phi_\om(u,\gamma) &=& \frac{u}{u-1}\int_{\gamma_0}^{\gamma}
d\gamma'\,\left[\frac{\om}{\alb u}-\chi(\gamma')\right]
\nonumber \\
&=& \frac{u}{u-1} 
\left[ \left( \frac{\om}{\alb u} + 2 \gE \right) (\gamma - \gamma_0) 
- \ln \frac{\Gamma(1-\gamma) \Gamma(\gamma_0)}{\Gamma(\gamma) 
\Gamma(1-\gamma_0)} \right]
\;,
\end{eqnarray}
$\chi$ being the Lipatov characteristic function, see 
eq.~(\ref{omlip}). To determine $\tilde g_\om(u,\gamma_0,T)$ we
make use of the boundary condition provided by eq.~(\ref{gomu}),
\begin{equation}\label{gomuT}
g_\om(u,T,T) = \frac{u\alb}{\om+\alb T}\left[1+g_\om t+
2\sum_{m=1}^\infty\zeta(2m+1)\,
\frac{\partial^{2m}g_\om}{\partial t^{2m}}\right]_{t=T}\;,
\end{equation}
which tells us that
\begin{equation}\label{gomubc}
\frac{1}{2\pi i}\tilde g_\om(u,\gamma_0,T)\int_\Gamma d\gamma
\left[\frac{\om}{\alb u}-\frac{u-1}{u}T
+\frac 1\gamma-\chi(\gamma)\right]
\,e^{-\gamma T+\phi_\om(u,\gamma)} =1\;.
\end{equation}
Deleting a total derivative from the integrand, since endpoint contributions
must vanish, we obtain simply
\begin{equation}\label{gomu0}
\frac{1}{2\pi i}\tilde g_\om(u,\gamma_0,T)
=\left[I_\om(u,T)\right]^{-1}\,,
\end{equation}
where
\begin{equation}\label{Idef}
I_\om(u,T) =\int_\Gamma \frac{d\gamma}{\gamma}
\,e^{-\gamma T+\phi_\om(u,\gamma)}\,.
\end{equation}
Hence
\begin{equation}\label{gomsol}
g_\om(u,t,T) =\frac{1}{I_\om(u,T)}
\int_\Gamma d\gamma\,e^{-\gamma t+\phi_\om(u,\gamma)}
\end{equation}
so that finally
\begin{equation}\label{Gomsol}
G_\om(u,T)= 1+\int_0^T dt\,g_\om(u,t,T)=\frac{I_\om(u,0)}{I_\om(u,T)}\;.
\end{equation}

\section{Jet multiplicity moments}\label{sec_mom}
The moments of the jet multiplicity distribution are obtained by
successive differentiation of the generating function at $u=1$:
\begin{equation}\label{rmoms}
\VEV{r(r-1)\ldots(r-s+1)}_\om =
\left.\frac{\partial^s R_\om}{\partial u^s}\right|_{u=1}
=\exp[-\glip(\alb/\om)T]
\left.\frac{\partial^s G_\om}{\partial u^s}\right|_{u=1}\;.
\end{equation}
To compute these quantities to all orders from the expression (\ref{Gomsol})
we can use the method of steepest descent. It is clear
from eq.~(\ref{phidef}) that the saddle point of
$\phi_\om(u,\gamma)$ occurs at $\gamma=\gamma_s$ where
$\om=\alb u\chi(\gamma_s)$, or in other words, from  eq.~(\ref{omlip}),
$\gamma_s=\glip(\alb u/\om)$.  Choosing for convenience
$\gamma_0=\gamma_s$, we have
\begin{equation}\label{phiexp}
\phi_\om(u,\gamma) = -\frac{u}{u-1}\sum_{n=2}^\infty \frac{1}{n!}
(\gamma-\gamma_s)^n \chi^{(n-1)}(\gamma_s)\;.
\end{equation}
Now we have to evaluate integrals of the form\footnote{The saddle-point 
method has been applied to similar integrals in ref.~\protect\cite{CCF}.} 
\begin{eqnarray}\label{intexp}
&&\int_\Gamma d\gamma\,f(\gamma)\,e^{\phi_\om(u,\gamma)}
=\int_\Gamma d\gamma\sum_{m=0}^\infty\frac{f^{(m)}_s}{m!}
(\gamma-\gamma_s)^m\,e^{\phi_\om(u,\gamma)}\nonumber\\
&&=\Biggl\{f_s+\frac{(u-1)}{u\chi'_s}\left(\frac 12 f''_s
-\frac 12\frac{\chi''_s}{\chi'_s}f'_s\right)\nonumber\\
&&+\left(\frac{u-1}{u\chi'_s}\right)^2
\Biggl[\frac 18 f''''_s-\frac 5{12}\frac{\chi''_s}{\chi'_s}f'''_s
-\frac 14\frac{\chi'''_s}{\chi'_s}f''_s
+\frac 58\frac{\chi''^2_s}{\chi'^2_s}f''_s\nonumber\\
&&-\frac 18\frac{\chi''''_s}{\chi'_s}f'_s
+\frac 23\frac{\chi''_s\chi'''_s}{\chi'^2_s}f'_s
-\frac 58\left(\frac{\chi''_s}{\chi'_s}\right)^3 f'_s\Biggr]
+\ldots\Biggr\}
\int_\Gamma d\gamma\,e^{\phi_\om(u,\gamma)}
\end{eqnarray}
where $f_s=f(\gamma_s)$, $\chi'_s=\chi'(\gamma_s)$, etc., and the
dots represent terms of higher order in $(u-1)$.  To express everything
in terms of $f(\glip)$, $\chi'(\glip)$, etc., where
$\glip\equiv\glip(\alb/\om)$, we use the results in the Appendix to write
\begin{eqnarray}\label{gamsexp}
\gamma_s &=& \glip +\glip'\,(u-1)\alom
+ \frac 12\glip''\,(u-1)^2\left(\alom\right)^2+\ldots\nonumber\\
&=& \glip -(u-1)\frac{\chi}{\chi'}+(u-1)^2
\left(\frac{\chi}{\chi'}-\frac{\chi^2\chi''}{2\chi'^3}\right)+\ldots
\end{eqnarray}
where now $\chi=\chi(\glip)$, etc., and we can use this to 
expand eq.~(\ref{intexp}). 

In the present case we have
\begin{equation}\label{fgam}
f(\gamma) = \frac{1}{\gamma}e^{-\gamma T}\;.
\end{equation}
Hence from eqs.~(\ref{Gomsol}), (\ref{rmoms}) and (\ref{intexp}) we obtain
for the mean number of jets,
\begin{equation}\label{meanres}
\VEV{r}_\om = 
- \frac{1}{\chi'} 
\left(\frac{1}{\glip} + \frac{\chi''}{2 \chi'}+ \chi \right) T
- \frac{1}{2 \chi'} T^2
\end{equation}
and for the variance
$\sig^2_\om\equiv\VEV{r^2}_\om -\VEV{r}^2_\om$,
\begin{eqnarray}\label{varres}
\sig^2_\om&=& 
\frac{1}{\chi'} 
  \left[ \frac{1}{\glip} + \chi
         - \frac{4}{\glip^3 \chi'}
         - \frac{2 \chi}{\glip^2 \chi'}
         - \frac{\chi^2 \chi''}{\chi'^2}
         - \frac{3 \chi''}{\glip^2 \chi'^2}
  \right. \nonumber \\
&&\hspace{1cm} \left.
         - \frac{2 \chi \chi''}{\glip \chi'^2}
         + \frac{\chi''}{2 \chi'} 
         - \frac{2 \chi \chi''^2}{\chi'^3}
         - \frac{2 \chi''^2}{\glip \chi'^3}
         - \frac{5 \chi''^3}{4 \chi'^4}
  \right. \nonumber \\
&&\hspace{1cm} \left.
         + \frac{\chi \chi'''}{\chi'^2}
         + \frac{\chi'''}{\glip \chi'^2}
         + \frac{4 \chi'' \chi'''}{3 \chi'^3}
         - \frac{\chi''''}{4 \chi'^2}
  \right] T
\nonumber \\
&& + \frac{1}{\chi'} 
   \left( \frac{1}{2}
          - \frac{1}{\glip^2 \chi'} 
          - \frac{\chi \chi''}{\chi'^2}
          - \frac{\chi''}{\glip \chi'^2}
          - \frac{\chi''^2}{\chi'^3}
          + \frac{\chi'''}{2 \chi'^2}
   \right) T^2
\nonumber \\
&& - \frac{\chi''}{3 \chi'^3} T^3\;.
\end{eqnarray}

Eqs.~(\ref{meanres}) and (\ref{varres}) are convenient for numerical
calculation, since they only involve the evaluation of polygamma functions.
To compare with perturbative expansions, we can use the results in the
Appendix to rewrite them in terms of $\glip$ and its derivatives:
\begin{eqnarray} \label{meanrespert}
\VEV{r}_\om &=&
\left[
-1 + \glip' + \alom \left(\frac{\glip'}{\glip} 
- \frac{\glip''}{2\glip'} 
\right) \right]\alom T
+ \frac{1}{2}\glip' \left( \alom \right)^2 T^2\;,\\
\sig^2_\om &=&
  \left[
    -1 + \glip' 
    + \alom \left( 
             - \frac{2}{\glip} 
             + \frac{3 \glip'}{\glip}
             + \glip'' 
             - \frac{3 \glip''}{2 \glip'} 
            \right)  
  \right. \nonumber \\
&&\hspace{0.2cm} \left.
    + \left(\alom\right)^2 
            \left( 
             \frac{6 \glip'}{\glip^2} 
             - \frac{2 \glip'^2}{\glip^2} 
             + \frac{2 \glip''}{\glip} 
             - \frac{2 \glip''}{\glip \glip'} 
             - \frac{\glip''^2}{\glip'^3} 
             + \frac{\glip''^2}{\glip'^2} 
             + \frac{2 \glip'''}{3 \glip'^2} 
             - \frac{\glip'''}{\glip'} 
            \right)  
  \right. \nonumber \\
&&\hspace{0.2cm} \left.
    + \left(\alom\right)^3 
            \left( 
             \frac{-4 \glip'^2}{\glip^3} 
             + \frac{3 \glip''}{\glip^2} 
             + \frac{\glip''^2}{\glip \glip'^2} 
             + \frac{\glip''^3}{\glip'^4} 
             - \frac{\glip'''}{\glip \glip'} 
             - \frac{7 \glip'' \glip'''}{6 \glip'^3} 
             + \frac{\glip''''}{4 \glip'^2} 
      \right)
  \right] \alom T \nonumber \\
&&+\left[ -1 + \frac{3 \glip'}{2} 
          + \alom \left( \glip'' 
                         + \frac{2 \glip'}{\glip} 
                         - \frac{\glip''}{\glip'} 
                  \right) 
  \right. \nonumber \\
&&\hspace{1cm} \left.
          + \left(\alom\right)^2 
              \left(- \frac{\glip'^2}{\glip^2} 
                    + \frac{\glip''}{\glip} 
                    + \frac{\glip''^2}{2 \glip'^2} 
                    - \frac{\glip'''}{2 \glip'} 
              \right)
\right] \left(\alom\right)^2 T^2 \nonumber \\
&&+\frac{1}{3} \left( 
2 \glip' + \alom \glip''
\right) \left(\alom\right)^3 T^3
\end{eqnarray}
Using the perturbative expansion (\ref{glip})  for $\glip$, we find
\begin{eqnarray}\label{PTmean}
\VEV{r}_\om &=&
\alom T +\frac{1}{2}\left(\alom\right)^2 T^2
+2\zeta(3)\left(\alom\right)^4 T\nonumber\\
&&+4\zeta(3)\left(\alom\right)^5 T^2
-8\zeta(5)\left(\alom\right)^6 T +\cdots
\end{eqnarray}
and
\begin{eqnarray}\label{PTvar}
\sig^2_\om &=&
\alom T+\frac{3}{2}\left(\alom\right)^2 T^2
+\frac{2}{3}\left(\alom\right)^3 T^3
-2\zeta(3)\left(\alom\right)^4 T\nonumber\\
&&+12\zeta(3)\left(\alom\right)^5 T^2
-\left(8\zeta(5) -\frac{40}{3}\zeta(3)T^2\right)
\left(\alom\right)^6 T+\cdots\;,
\end{eqnarray}
which do indeed agree with the expansions obtained in ref.~\cite{EW1}.

\section{Discussion of the jet multiplicity moments}\label{sec_disc}
We see from eqs.~(\ref{meanres}) and (\ref{varres}) that it
remains true to all orders to SL precision that, as in the DL
approximation \cite{Web98}, the mean number of jets is a quadratic
function of $T$ while the variance is a cubic function of $T$.
Thus the distribution of jet multiplicity at small $x$ and large
$T$ remains narrow, in the sense that its r.m.s.\ width
increases less rapidly than its mean with increasing $T$.

It is interesting to see how rapidly the perturbative expansions for the 
mean number of mini-jets and its variance 
converge. We first introduce the notation 
\begin{equation}\label{coeffr}
\VEV{r}_\om = a_1 \left( \alom \right) T
+ a_2 \left( \alom \right)^2 T^2
\end{equation}
and 
\begin{equation} \label{coeffr2}
\sig^2_\om=
b_1 \left( \alom \right) T + b_2 \left( \alom \right)^2 T^2
+ b_3 \left( \alom \right)^3 T^3
\,.
\end{equation}
Fig.\ \ref{fig_cor} shows the coefficients 
$a_i$ ($i=1,2$) as functions of $\alb/\om$, including the 
exact dependence according to eq.~(\ref{meanres}) as well 
as the perturbative expressions (\ref{meanrespert}) 
expanded to different orders. 
The coefficients $b_i$ ($i=1,2,3$) are shown in 
fig.\ \ref{fig_cor2}. %
\FIGURE[ht]{\begin{picture}(0,0)%
\epsfig{file=cor.pstex}%
\end{picture}%
\setlength{\unitlength}{4144sp}%
\begingroup\makeatletter\ifx\SetFigFont\undefined%
\gdef\SetFigFont#1#2#3#4#5{%
  \reset@font\fontsize{#1}{#2pt}%
  \fontfamily{#3}\fontseries{#4}\fontshape{#5}%
  \selectfont}%
\fi\endgroup%
\begin{picture}(6074,2976)(331,-2861)
\put(639,-2462){\makebox(0,0)[rb]{\smash{\SetFigFont{10}{12.0}{\familydefault}{\mddefault}{\updefault}-0.5}}}
\put(639,-1844){\makebox(0,0)[rb]{\smash{\SetFigFont{10}{12.0}{\familydefault}{\mddefault}{\updefault}0}}}
\put(639,-1226){\makebox(0,0)[rb]{\smash{\SetFigFont{10}{12.0}{\familydefault}{\mddefault}{\updefault}0.5}}}
\put(639,-608){\makebox(0,0)[rb]{\smash{\SetFigFont{10}{12.0}{\familydefault}{\mddefault}{\updefault}1}}}
\put(639, 10){\makebox(0,0)[rb]{\smash{\SetFigFont{10}{12.0}{\familydefault}{\mddefault}{\updefault}1.5}}}
\put(709,-2580){\makebox(0,0)[b]{\smash{\SetFigFont{10}{12.0}{\familydefault}{\mddefault}{\updefault}0}}}
\put(1327,-2580){\makebox(0,0)[b]{\smash{\SetFigFont{10}{12.0}{\familydefault}{\mddefault}{\updefault}0.1}}}
\put(1946,-2580){\makebox(0,0)[b]{\smash{\SetFigFont{10}{12.0}{\familydefault}{\mddefault}{\updefault}0.2}}}
\put(2563,-2580){\makebox(0,0)[b]{\smash{\SetFigFont{10}{12.0}{\familydefault}{\mddefault}{\updefault}0.3}}}
\put(3181,-2580){\makebox(0,0)[b]{\smash{\SetFigFont{10}{12.0}{\familydefault}{\mddefault}{\updefault}0.4}}}
\put(1945,-2816){\makebox(0,0)[b]{\smash{\SetFigFont{10}{12.0}{\familydefault}{\mddefault}{\updefault}$\alb/\omega$}}}
\put(1018,-299){\makebox(0,0)[lb]{\smash{\SetFigFont{10}{12.0}{\familydefault}{\mddefault}{\updefault}$a_1$}}}
\put(1813,-1597){\makebox(0,0)[rb]{\smash{\SetFigFont{10}{12.0}{\familydefault}{\mddefault}{\updefault}exact}}}
\put(1813,-1757){\makebox(0,0)[rb]{\smash{\SetFigFont{10}{12.0}{\familydefault}{\mddefault}{\updefault}$10^{\mbox{\scriptsize th}}$ order}}}
\put(1813,-1917){\makebox(0,0)[rb]{\smash{\SetFigFont{10}{12.0}{\familydefault}{\mddefault}{\updefault}$14^{\mbox{\scriptsize th}}$ order}}}
\put(1813,-2077){\makebox(0,0)[rb]{\smash{\SetFigFont{10}{12.0}{\familydefault}{\mddefault}{\updefault}$15^{\mbox{\scriptsize th}}$ order}}}
\put(3763,-2462){\makebox(0,0)[rb]{\smash{\SetFigFont{10}{12.0}{\familydefault}{\mddefault}{\updefault}0}}}
\put(3763,-1844){\makebox(0,0)[rb]{\smash{\SetFigFont{10}{12.0}{\familydefault}{\mddefault}{\updefault}0.5}}}
\put(3763,-1226){\makebox(0,0)[rb]{\smash{\SetFigFont{10}{12.0}{\familydefault}{\mddefault}{\updefault}1}}}
\put(3763,-608){\makebox(0,0)[rb]{\smash{\SetFigFont{10}{12.0}{\familydefault}{\mddefault}{\updefault}1.5}}}
\put(3763, 10){\makebox(0,0)[rb]{\smash{\SetFigFont{10}{12.0}{\familydefault}{\mddefault}{\updefault}2}}}
\put(3834,-2580){\makebox(0,0)[b]{\smash{\SetFigFont{10}{12.0}{\familydefault}{\mddefault}{\updefault}0}}}
\put(4452,-2580){\makebox(0,0)[b]{\smash{\SetFigFont{10}{12.0}{\familydefault}{\mddefault}{\updefault}0.1}}}
\put(5070,-2580){\makebox(0,0)[b]{\smash{\SetFigFont{10}{12.0}{\familydefault}{\mddefault}{\updefault}0.2}}}
\put(5687,-2580){\makebox(0,0)[b]{\smash{\SetFigFont{10}{12.0}{\familydefault}{\mddefault}{\updefault}0.3}}}
\put(6305,-2580){\makebox(0,0)[b]{\smash{\SetFigFont{10}{12.0}{\familydefault}{\mddefault}{\updefault}0.4}}}
\put(5069,-2816){\makebox(0,0)[b]{\smash{\SetFigFont{10}{12.0}{\familydefault}{\mddefault}{\updefault}$\alb/\omega$}}}
\put(4142,-299){\makebox(0,0)[lb]{\smash{\SetFigFont{10}{12.0}{\familydefault}{\mddefault}{\updefault}$a_2$}}}
\end{picture}

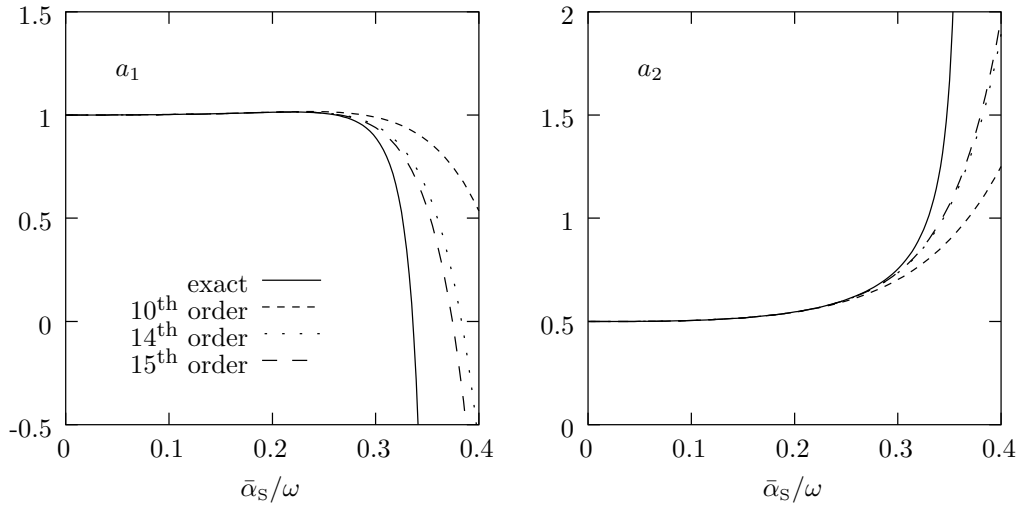
  \caption{Coefficients in the mean jet multiplicity 
    $\VEV{r}_\om$ as functions 
    of $\alb/\om$, as defined in (\protect\ref{coeffr}). 
    The solid line represents the exact coefficients as calculated from 
    (\protect\ref{meanres}), the other lines are obtained by 
    expanding (\protect\ref{meanrespert}) to 10th, 14th, and 
    15th order in $\alb/\om$.}
\label{fig_cor}
}
\FIGURE[ht]{\begin{picture}(0,0)%
\epsfig{file=cor2.pstex}%
\end{picture}%
\setlength{\unitlength}{4144sp}%
\begingroup\makeatletter\ifx\SetFigFont\undefined%
\gdef\SetFigFont#1#2#3#4#5{%
  \reset@font\fontsize{#1}{#2pt}%
  \fontfamily{#3}\fontseries{#4}\fontshape{#5}%
  \selectfont}%
\fi\endgroup%
\begin{picture}(5992,6223)(413,-6108)
\put(613,-2462){\makebox(0,0)[rb]{\smash{\SetFigFont{10}{12.0}{\familydefault}{\mddefault}{\updefault}0.5}}}
\put(613,-1844){\makebox(0,0)[rb]{\smash{\SetFigFont{10}{12.0}{\familydefault}{\mddefault}{\updefault}1}}}
\put(613,-1226){\makebox(0,0)[rb]{\smash{\SetFigFont{10}{12.0}{\familydefault}{\mddefault}{\updefault}1.5}}}
\put(613,-608){\makebox(0,0)[rb]{\smash{\SetFigFont{10}{12.0}{\familydefault}{\mddefault}{\updefault}2}}}
\put(613, 10){\makebox(0,0)[rb]{\smash{\SetFigFont{10}{12.0}{\familydefault}{\mddefault}{\updefault}2.5}}}
\put(684,-2580){\makebox(0,0)[b]{\smash{\SetFigFont{10}{12.0}{\familydefault}{\mddefault}{\updefault}0}}}
\put(1302,-2580){\makebox(0,0)[b]{\smash{\SetFigFont{10}{12.0}{\familydefault}{\mddefault}{\updefault}0.1}}}
\put(1920,-2580){\makebox(0,0)[b]{\smash{\SetFigFont{10}{12.0}{\familydefault}{\mddefault}{\updefault}0.2}}}
\put(2537,-2580){\makebox(0,0)[b]{\smash{\SetFigFont{10}{12.0}{\familydefault}{\mddefault}{\updefault}0.3}}}
\put(3155,-2580){\makebox(0,0)[b]{\smash{\SetFigFont{10}{12.0}{\familydefault}{\mddefault}{\updefault}0.4}}}
\put(1919,-2816){\makebox(0,0)[b]{\smash{\SetFigFont{10}{12.0}{\familydefault}{\mddefault}{\updefault}$\alb/\omega$}}}
\put(992,-299){\makebox(0,0)[lb]{\smash{\SetFigFont{10}{12.0}{\familydefault}{\mddefault}{\updefault}$b_1$}}}
\put(2191,-5708){\makebox(0,0)[rb]{\smash{\SetFigFont{10}{12.0}{\familydefault}{\mddefault}{\updefault}0.5}}}
\put(2191,-5090){\makebox(0,0)[rb]{\smash{\SetFigFont{10}{12.0}{\familydefault}{\mddefault}{\updefault}1}}}
\put(2191,-4472){\makebox(0,0)[rb]{\smash{\SetFigFont{10}{12.0}{\familydefault}{\mddefault}{\updefault}1.5}}}
\put(2191,-3855){\makebox(0,0)[rb]{\smash{\SetFigFont{10}{12.0}{\familydefault}{\mddefault}{\updefault}2}}}
\put(2191,-3237){\makebox(0,0)[rb]{\smash{\SetFigFont{10}{12.0}{\familydefault}{\mddefault}{\updefault}2.5}}}
\put(2262,-5827){\makebox(0,0)[b]{\smash{\SetFigFont{10}{12.0}{\familydefault}{\mddefault}{\updefault}0}}}
\put(2880,-5827){\makebox(0,0)[b]{\smash{\SetFigFont{10}{12.0}{\familydefault}{\mddefault}{\updefault}0.1}}}
\put(3498,-5827){\makebox(0,0)[b]{\smash{\SetFigFont{10}{12.0}{\familydefault}{\mddefault}{\updefault}0.2}}}
\put(4115,-5827){\makebox(0,0)[b]{\smash{\SetFigFont{10}{12.0}{\familydefault}{\mddefault}{\updefault}0.3}}}
\put(4733,-5827){\makebox(0,0)[b]{\smash{\SetFigFont{10}{12.0}{\familydefault}{\mddefault}{\updefault}0.4}}}
\put(3497,-6063){\makebox(0,0)[b]{\smash{\SetFigFont{10}{12.0}{\familydefault}{\mddefault}{\updefault}$\alb/\omega$}}}
\put(2570,-3545){\makebox(0,0)[lb]{\smash{\SetFigFont{10}{12.0}{\familydefault}{\mddefault}{\updefault}$b_3$}}}
\put(3763,-1844){\makebox(0,0)[rb]{\smash{\SetFigFont{10}{12.0}{\familydefault}{\mddefault}{\updefault}0.5}}}
\put(3763,-1226){\makebox(0,0)[rb]{\smash{\SetFigFont{10}{12.0}{\familydefault}{\mddefault}{\updefault}1}}}
\put(3763,-608){\makebox(0,0)[rb]{\smash{\SetFigFont{10}{12.0}{\familydefault}{\mddefault}{\updefault}1.5}}}
\put(3763, 10){\makebox(0,0)[rb]{\smash{\SetFigFont{10}{12.0}{\familydefault}{\mddefault}{\updefault}2}}}
\put(3834,-2580){\makebox(0,0)[b]{\smash{\SetFigFont{10}{12.0}{\familydefault}{\mddefault}{\updefault}0}}}
\put(4452,-2580){\makebox(0,0)[b]{\smash{\SetFigFont{10}{12.0}{\familydefault}{\mddefault}{\updefault}0.1}}}
\put(5070,-2580){\makebox(0,0)[b]{\smash{\SetFigFont{10}{12.0}{\familydefault}{\mddefault}{\updefault}0.2}}}
\put(5687,-2580){\makebox(0,0)[b]{\smash{\SetFigFont{10}{12.0}{\familydefault}{\mddefault}{\updefault}0.3}}}
\put(6305,-2580){\makebox(0,0)[b]{\smash{\SetFigFont{10}{12.0}{\familydefault}{\mddefault}{\updefault}0.4}}}
\put(5069,-2816){\makebox(0,0)[b]{\smash{\SetFigFont{10}{12.0}{\familydefault}{\mddefault}{\updefault}$\alb/\omega$}}}
\put(4142,-299){\makebox(0,0)[lb]{\smash{\SetFigFont{10}{12.0}{\familydefault}{\mddefault}{\updefault}$b_2$}}}
\put(3763,-2462){\makebox(0,0)[rb]{\smash{\SetFigFont{10}{12.0}{\familydefault}{\mddefault}{\updefault}0}}}
\end{picture}

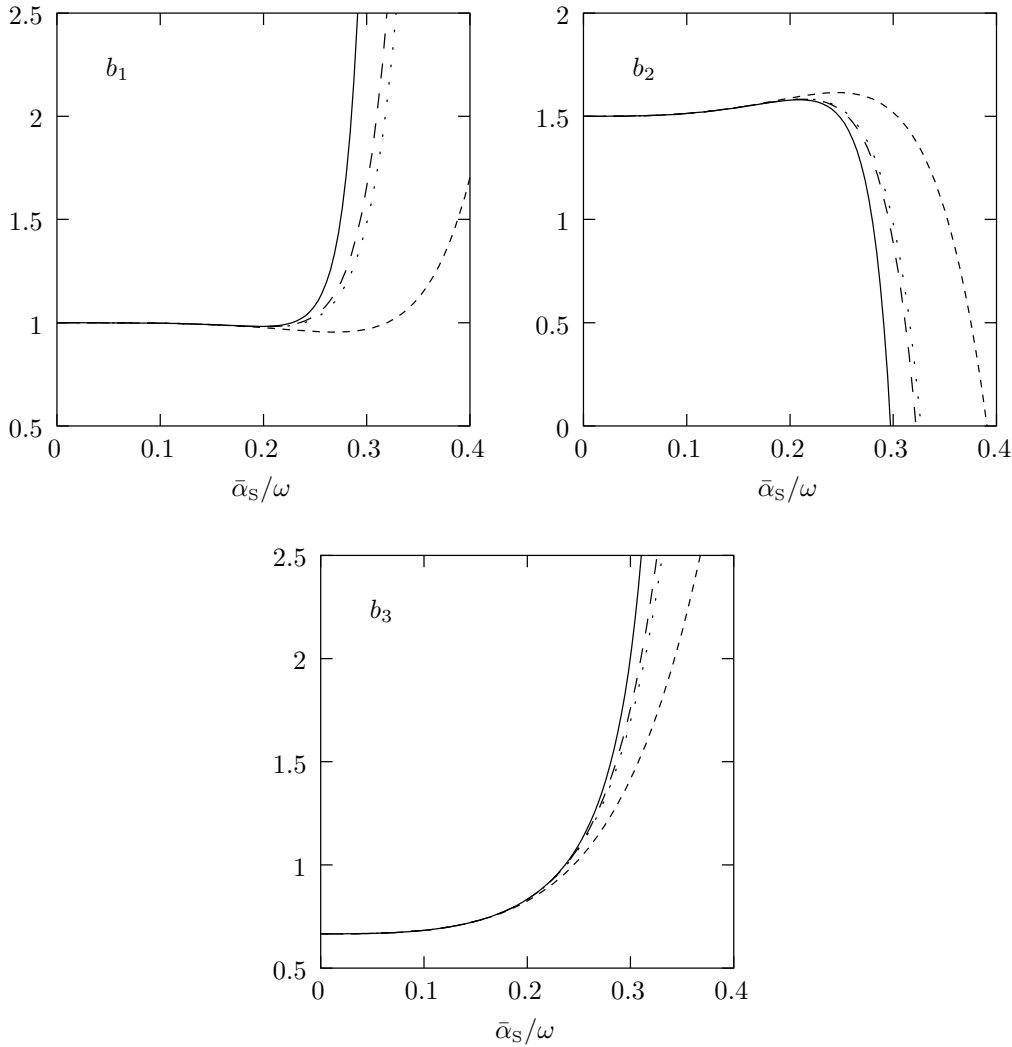
  \caption{Coefficients in the variance $\sig^2_\om$ as functions 
    of $\alb/\om$, as defined in 
    (\protect\ref{coeffr2}). Lines as in fig.\ \protect\ref{fig_cor}.}
\label{fig_cor2}
}

The coefficients in the mean jet multiplicity $\VEV{r}_\om$ are 
extremely well approximated by the perturbative expressions 
for small values of $\alb/\om$, up to $0.3$. In this region, the
coefficients remain close to the (constant) values predicted
by the double-logarithmic approximation.
Above $\alb/\om \simeq 0.3$,  where the coefficients start to diverge,
the perturbative expansion converges rather slowly. 
A similar behaviour is observed for the coefficients in the variance 
$\sig^2_\om$. The perturbative expansion converges very 
rapidly for small $\alb/\om$, in this case up to $\sim 0.25$. 
Above $\alb/\om \simeq 0.3$, the convergence is again rather poor. 

The singularities in the coefficients arise from the fact that
eq.~(\ref{omlip}) does not have real solutions for
$\alb/\om >(4 \ln 2)^{-1} = 0.361$. 
This bound will be raised \cite{CCF} when NLO corrections 
to the BFKL equation \cite{NLOA,NLOB} are included. 
We therefore expect higher order corrections to the BFKL equation 
to have a significant effect on the jet multiplicity moments for 
values of $\alb/\om$ above $\sim 0.3$. 

For practical applications, we need to compute the multiplicity moments as
functions of $x$.  This is done most conveniently using the perturbative
expansions (\ref{PTmean}) and (\ref{PTvar}), extended to sufficiently
high order, and performing the inverse Mellin transformation term
by term using 
\begin{equation}\label{melom}
\frac{1}{2\pi i}\int_C d\om\,x^{-\om-1}\left(\alom\right)^n
=\frac{\alb}{x}\frac{[\alb\ln(1/x)]^{n-1}}{(n-1)!}\;.
\end{equation}
The factorial suppression of high orders in $x$-space removes the
convergence problems that were encountered in $\om$-space.

To illustrate the behaviour in $x$-space, we consider the
mini-jet multiplicity associated with pointlike scattering on
the gluonic component of the proton at small $x$.  The mean
number of mini-jets considered as a function of $x$ is given by
\begin{equation} \label{nx}
n(x)=\frac{F(x,Q^2) \otimes \VEV{r}(x)}{F(x,Q^2)} \,.
\end{equation}
where $\VEV{r}(x)$ is the inverse Mellin transform of $\VEV{r}_\om$,
obtained by means of eq.~(\ref{melom}). 
Similarly the dispersion $\sigma_n$ in the mean 
number of mini-jets 
as a function of $x$ is 
\begin{equation} \label{sigx}
\sigma_n^2(x) = \frac{F(x,Q^2) \otimes \VEV{r^2}(x)}{F(x,Q^2)}
 - [n(x)]^2 \,.
\end{equation}

\FIGURE[ht]{
\begin{picture}(0,0)%
\epsfig{file=disres1.pstex}%
\end{picture}%
\setlength{\unitlength}{3947sp}%
\begingroup\makeatletter\ifx\SetFigFont\undefined%
\gdef\SetFigFont#1#2#3#4#5{%
  \reset@font\fontsize{#1}{#2pt}%
  \fontfamily{#3}\fontseries{#4}\fontshape{#5}%
  \selectfont}%
\fi\endgroup%
\begin{picture}(6826,3072)(1037,-2999)
\put(4094,-2751){\makebox(0,0)[b]{\smash{\SetFigFont{10}{12.0}{\familydefault}{\mddefault}{\updefault}$10^{-1}$}}}
\put(2992,-959){\makebox(0,0)[lb]{\smash{\SetFigFont{10}{12.0}{\familydefault}{\mddefault}{\updefault}$\mR\!=\!1\,\mbox{GeV}$}}}
\put(1425,-2256){\makebox(0,0)[rb]{\smash{\SetFigFont{10}{12.0}{\familydefault}{\mddefault}{\updefault}1}}}
\put(1425,-1886){\makebox(0,0)[rb]{\smash{\SetFigFont{10}{12.0}{\familydefault}{\mddefault}{\updefault}2}}}
\put(1425,-1515){\makebox(0,0)[rb]{\smash{\SetFigFont{10}{12.0}{\familydefault}{\mddefault}{\updefault}3}}}
\put(1425,-1144){\makebox(0,0)[rb]{\smash{\SetFigFont{10}{12.0}{\familydefault}{\mddefault}{\updefault}4}}}
\put(1425,-773){\makebox(0,0)[rb]{\smash{\SetFigFont{10}{12.0}{\familydefault}{\mddefault}{\updefault}5}}}
\put(1425,-403){\makebox(0,0)[rb]{\smash{\SetFigFont{10}{12.0}{\familydefault}{\mddefault}{\updefault}6}}}
\put(1425,-32){\makebox(0,0)[rb]{\smash{\SetFigFont{10}{12.0}{\familydefault}{\mddefault}{\updefault}7}}}
\put(1499,-2751){\makebox(0,0)[b]{\smash{\SetFigFont{10}{12.0}{\familydefault}{\mddefault}{\updefault}$10^{-5}$}}}
\put(2148,-2751){\makebox(0,0)[b]{\smash{\SetFigFont{10}{12.0}{\familydefault}{\mddefault}{\updefault}$10^{-4}$}}}
\put(2796,-2751){\makebox(0,0)[b]{\smash{\SetFigFont{10}{12.0}{\familydefault}{\mddefault}{\updefault}$10^{-3}$}}}
\put(3445,-2751){\makebox(0,0)[b]{\smash{\SetFigFont{10}{12.0}{\familydefault}{\mddefault}{\updefault}$10^{-2}$}}}
\put(1153,-1330){\makebox(0,0)[b]{\smash{\SetFigFont{10}{12.0}{\familydefault}{\mddefault}{\updefault}$n$}}}
\put(2796,-2999){\makebox(0,0)[b]{\smash{\SetFigFont{10}{12.0}{\familydefault}{\mddefault}{\updefault}$x$}}}
\put(2457,-2256){\makebox(0,0)[lb]{\smash{\SetFigFont{10}{12.0}{\familydefault}{\mddefault}{\updefault}$\mR\!=\!5\,\mbox{GeV}$}}}
\put(4848,-2256){\makebox(0,0)[rb]{\smash{\SetFigFont{10}{12.0}{\familydefault}{\mddefault}{\updefault}1}}}
\put(4848,-1886){\makebox(0,0)[rb]{\smash{\SetFigFont{10}{12.0}{\familydefault}{\mddefault}{\updefault}2}}}
\put(4848,-1515){\makebox(0,0)[rb]{\smash{\SetFigFont{10}{12.0}{\familydefault}{\mddefault}{\updefault}3}}}
\put(4848,-1144){\makebox(0,0)[rb]{\smash{\SetFigFont{10}{12.0}{\familydefault}{\mddefault}{\updefault}4}}}
\put(4848,-773){\makebox(0,0)[rb]{\smash{\SetFigFont{10}{12.0}{\familydefault}{\mddefault}{\updefault}5}}}
\put(4848,-403){\makebox(0,0)[rb]{\smash{\SetFigFont{10}{12.0}{\familydefault}{\mddefault}{\updefault}6}}}
\put(4848,-32){\makebox(0,0)[rb]{\smash{\SetFigFont{10}{12.0}{\familydefault}{\mddefault}{\updefault}7}}}
\put(4922,-2751){\makebox(0,0)[b]{\smash{\SetFigFont{10}{12.0}{\familydefault}{\mddefault}{\updefault}$10^{-5}$}}}
\put(5571,-2751){\makebox(0,0)[b]{\smash{\SetFigFont{10}{12.0}{\familydefault}{\mddefault}{\updefault}$10^{-4}$}}}
\put(6219,-2751){\makebox(0,0)[b]{\smash{\SetFigFont{10}{12.0}{\familydefault}{\mddefault}{\updefault}$10^{-3}$}}}
\put(6868,-2751){\makebox(0,0)[b]{\smash{\SetFigFont{10}{12.0}{\familydefault}{\mddefault}{\updefault}$10^{-2}$}}}
\put(7517,-2751){\makebox(0,0)[b]{\smash{\SetFigFont{10}{12.0}{\familydefault}{\mddefault}{\updefault}$10^{-1}$}}}
\put(4576,-1330){\makebox(0,0)[b]{\smash{\SetFigFont{10}{12.0}{\familydefault}{\mddefault}{\updefault}$\sigma_n$}}}
\put(6219,-2999){\makebox(0,0)[b]{\smash{\SetFigFont{10}{12.0}{\familydefault}{\mddefault}{\updefault}$x$}}}
\put(5551,-2236){\makebox(0,0)[lb]{\smash{\SetFigFont{10}{12.0}{\familydefault}{\mddefault}{\updefault}$\mR\!=\!5\,\mbox{GeV}$}}}
\put(6226,-1336){\makebox(0,0)[lb]{\smash{\SetFigFont{10}{12.0}{\familydefault}{\mddefault}{\updefault}$\mR\!=\!1\,\mbox{GeV}$}}}
\end{picture}

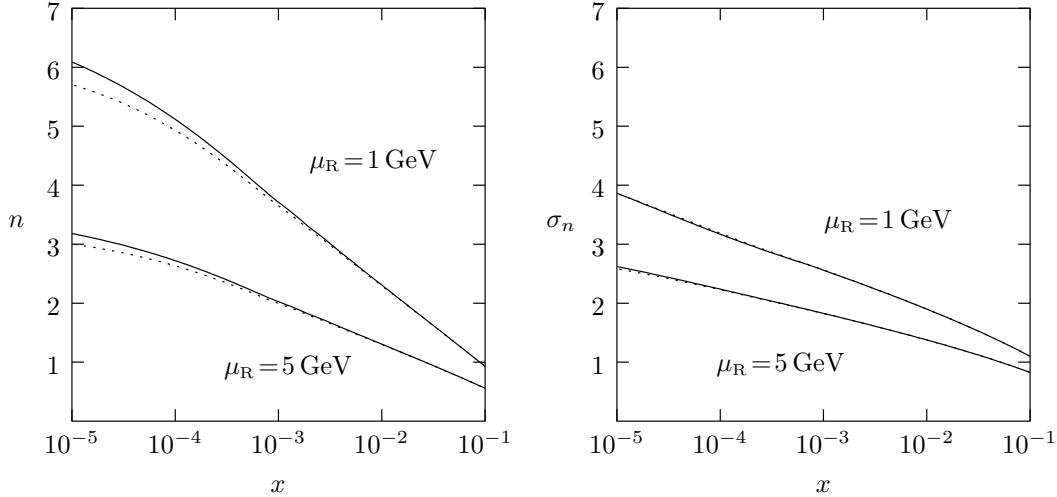
  \caption{The mean and dispersion of the mini-jet multiplicity
    as functions of $x$ at $Q\!=\!100\,\mbox{GeV}$ 
    for two different resolution scales $\mR$. The solid lines show the 
    SL results including the 15th order in perturbation theory, 
    the dashed lines correspond to the DL approximation.}
\label{fig_disjets}}

To perform the convolution in eqs.~(\ref{nx}), (\ref{sigx}) we used
the leading-order MRST gluon distribution \cite{MRST}.
Fig.~\ref{fig_disjets} shows the resulting $x$-dependence 
of the mean number of mini-jets $n(x)$ 
and its dispersion $\sigma_n(x)$ 
for $Q\!=\! 100\,\mbox{GeV}$ and two different 
values of the resolution scale $\mR$. We see that the results
remain close to the predictions of the double-logarithmic
approximation at all but the very smallest values of $x$
and $\mR$.  The convergence of the perturbation series
in $x$-space is so good that there is no visible difference
between the results 
at 5th and 15th order for the ranges of
$x$ and $\mR$ shown.

We should emphasise that the results in fig.~\ref{fig_disjets} are not
directly comparable with data on deep inelastic lepton scattering (DIS).
To make quantitative predictions for DIS one would need to perform
a further convolution with an impact factor that represents
the ``quark box'' coupling of the gluon to the virtual photon.
One should also take into account the production of mini-jets
from the quark box, together with non-perturbative effects such
as mini-jet emission from the proton remnant.

\section{Mini-jet multiplicity in Higgs production at LHC}\label{sec_higgs}

The dominant production process for a standard model Higgs boson 
at the LHC is expected to be gluon-gluon fusion. This process is 
illustrated in fig.~\ref{fig_higgs}. 
\FIGURE[h]{
\begin{picture}(0,0)%
\epsfig{file=higgs.pstex}%
\end{picture}%
\setlength{\unitlength}{4144sp}%
\begingroup\makeatletter\ifx\SetFigFont\undefined%
\gdef\SetFigFont#1#2#3#4#5{%
  \reset@font\fontsize{#1}{#2pt}%
  \fontfamily{#3}\fontseries{#4}\fontshape{#5}%
  \selectfont}%
\fi\endgroup%
\begin{picture}(3872,2094)(1800,-2233)
\put(2746,-466){\makebox(0,0)[lb]{\smash{\SetFigFont{12}{14.4}{\rmdefault}{\mddefault}{\updefault}$p_1$}}}
\put(2746,-1906){\makebox(0,0)[lb]{\smash{\SetFigFont{12}{14.4}{\rmdefault}{\mddefault}{\updefault}$p_2$}}}
\put(4051,-736){\makebox(0,0)[lb]{\smash{\SetFigFont{12}{14.4}{\rmdefault}{\mddefault}{\updefault}$\vdots$}}}
\put(4051,-1681){\makebox(0,0)[lb]{\smash{\SetFigFont{12}{14.4}{\rmdefault}{\mddefault}{\updefault}$\vdots$}}}
\put(4231,-781){\makebox(0,0)[lb]{\smash{\SetFigFont{12}{14.4}{\rmdefault}{\mddefault}{\updefault}${\Bigg \} }\,\,n_1$}}}
\put(4231,-1636){\makebox(0,0)[lb]{\smash{\SetFigFont{12}{14.4}{\rmdefault}{\mddefault}{\updefault}${\Bigg \} }\,\,n_2$}}}
\put(4546,-1276){\makebox(0,0)[lb]{\smash{\SetFigFont{12}{14.4}{\rmdefault}{\mddefault}{\updefault}$H$}}}
\put(3331,-1411){\makebox(0,0)[lb]{\smash{\SetFigFont{12}{14.4}{\rmdefault}{\mddefault}{\updefault}$x_2$}}}
\put(3331,-1051){\makebox(0,0)[lb]{\smash{\SetFigFont{12}{14.4}{\rmdefault}{\mddefault}{\updefault}$x_1$}}}
\end{picture}

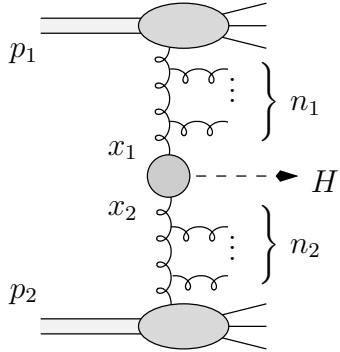
  \caption{Higgs boson production by gluon-gluon fusion.}
\label{fig_higgs}}
The production cross section for a Higgs boson of mass $M$ and rapidity $y$
by gluon-gluon fusion in proton-proton collisions at 
centre-of mass energy $\sqrt{s}$ takes the form
\cite{ESW}
\begin{equation}
\frac{d\sigma}{dy} = F(x_1,M^2)\,F(x_2,M^2)\,C(M^2)\,,
\end{equation}
where 
\begin{equation}
x_1 = \frac{M}{\sqrt{s}}\,e^y\;,\qquad\qquad
x_2 = \frac{M}{\sqrt{s}}\,e^{-y}\;,
\end{equation}
and for LHC $\sqrt{s}=14\,\mbox{TeV}$. 
$C$ represents the $gg\to H$ vertex, which is perturbatively 
calculable as an intermediate top-quark loop. In the 
mean number of mini-jets and its dispersion, however, 
$C$ cancels and we do not need its detailed form.
Thus we can compute the associated mini-jet multiplicity in Higgs
production quite simply using the machinery developed in the previous
section.

We will consider central production of the Higgs boson ($y=0$),
and therefore we have $x_1=x_2=x$ where $x=M/\sqrt{s}$. 
Since the gluon emissions in the upper and lower parts of 
fig.~\ref{fig_higgs} are independent, we can simply 
add the numbers $n_1=n(x_1)$ and $n_2=n(x_2)$ of mini-jets, 
and the mean multiplicity $N$ of associated mini-jets 
becomes\footnote{Again
we do not count any jets emerging from the proton remnants.} 
\begin{equation}
N(x) = n_1+ n_2 = 2 n(x) \,,
\end{equation}
where $n(x)$ can be calculated as in (\ref{nx}) after 
replacing $Q^2$ by $M^2$. 
Similarly, the variance is
\begin{equation}
\sigma^2_N(x) = \sigma^2_n(x_1)+\sigma^2_n(x_2)
= 2 \sigma^2_n(x)
\;.
\end{equation}
Again, $\sig^2_n$ can be obtained as in eq.~(\ref{sigx}). 

We have calculated the dependence of 
$N$ and $\sig_N$ on the Higgs mass $M$ and
our numerical results are shown in fig.~\ref{fig_higgsjets}. 
\FIGURE[h]{
\begin{picture}(0,0)%
\epsfig{file=higgsres1.pstex}%
\end{picture}%
\setlength{\unitlength}{3947sp}%
\begingroup\makeatletter\ifx\SetFigFont\undefined%
\gdef\SetFigFont#1#2#3#4#5{%
  \reset@font\fontsize{#1}{#2pt}%
  \fontfamily{#3}\fontseries{#4}\fontshape{#5}%
  \selectfont}%
\fi\endgroup%
\begin{picture}(6648,3072)(1061,-6899)
\put(1449,-6094){\makebox(0,0)[rb]{\smash{\SetFigFont{10}{12.0}{\familydefault}{\mddefault}{\updefault}1}}}
\put(1449,-5662){\makebox(0,0)[rb]{\smash{\SetFigFont{10}{12.0}{\familydefault}{\mddefault}{\updefault}2}}}
\put(1449,-5229){\makebox(0,0)[rb]{\smash{\SetFigFont{10}{12.0}{\familydefault}{\mddefault}{\updefault}3}}}
\put(1449,-4797){\makebox(0,0)[rb]{\smash{\SetFigFont{10}{12.0}{\familydefault}{\mddefault}{\updefault}4}}}
\put(1449,-4364){\makebox(0,0)[rb]{\smash{\SetFigFont{10}{12.0}{\familydefault}{\mddefault}{\updefault}5}}}
\put(1449,-3932){\makebox(0,0)[rb]{\smash{\SetFigFont{10}{12.0}{\familydefault}{\mddefault}{\updefault}6}}}
\put(1523,-6651){\makebox(0,0)[b]{\smash{\SetFigFont{10}{12.0}{\familydefault}{\mddefault}{\updefault}$70$}}}
\put(1994,-6651){\makebox(0,0)[b]{\smash{\SetFigFont{10}{12.0}{\familydefault}{\mddefault}{\updefault}$100$}}}
\put(2909,-6651){\makebox(0,0)[b]{\smash{\SetFigFont{10}{12.0}{\familydefault}{\mddefault}{\updefault}$200$}}}
\put(4118,-6651){\makebox(0,0)[b]{\smash{\SetFigFont{10}{12.0}{\familydefault}{\mddefault}{\updefault}$500$}}}
\put(1177,-5230){\makebox(0,0)[b]{\smash{\SetFigFont{10}{12.0}{\familydefault}{\mddefault}{\updefault}$N$}}}
\put(3203,-4191){\makebox(0,0)[lb]{\smash{\SetFigFont{10}{12.0}{\familydefault}{\mddefault}{\updefault}$\mR\!=\!1\,\mbox{GeV}$}}}
\put(2909,-5575){\makebox(0,0)[lb]{\smash{\SetFigFont{10}{12.0}{\familydefault}{\mddefault}{\updefault}$\mR\!=\!5\,\mbox{GeV}$}}}
\put(2820,-6899){\makebox(0,0)[b]{\smash{\SetFigFont{10}{12.0}{\familydefault}{\mddefault}{\updefault}$M\,[\mbox{GeV}]$}}}
\put(4848,-6094){\makebox(0,0)[rb]{\smash{\SetFigFont{10}{12.0}{\familydefault}{\mddefault}{\updefault}1}}}
\put(4848,-5662){\makebox(0,0)[rb]{\smash{\SetFigFont{10}{12.0}{\familydefault}{\mddefault}{\updefault}2}}}
\put(4848,-5229){\makebox(0,0)[rb]{\smash{\SetFigFont{10}{12.0}{\familydefault}{\mddefault}{\updefault}3}}}
\put(4848,-4797){\makebox(0,0)[rb]{\smash{\SetFigFont{10}{12.0}{\familydefault}{\mddefault}{\updefault}4}}}
\put(4848,-4364){\makebox(0,0)[rb]{\smash{\SetFigFont{10}{12.0}{\familydefault}{\mddefault}{\updefault}5}}}
\put(4848,-3932){\makebox(0,0)[rb]{\smash{\SetFigFont{10}{12.0}{\familydefault}{\mddefault}{\updefault}6}}}
\put(4922,-6651){\makebox(0,0)[b]{\smash{\SetFigFont{10}{12.0}{\familydefault}{\mddefault}{\updefault}$70$}}}
\put(5393,-6651){\makebox(0,0)[b]{\smash{\SetFigFont{10}{12.0}{\familydefault}{\mddefault}{\updefault}$100$}}}
\put(6308,-6651){\makebox(0,0)[b]{\smash{\SetFigFont{10}{12.0}{\familydefault}{\mddefault}{\updefault}$200$}}}
\put(7517,-6651){\makebox(0,0)[b]{\smash{\SetFigFont{10}{12.0}{\familydefault}{\mddefault}{\updefault}$500$}}}
\put(4576,-5230){\makebox(0,0)[b]{\smash{\SetFigFont{10}{12.0}{\familydefault}{\mddefault}{\updefault}$\sigma_N$}}}
\put(6219,-6899){\makebox(0,0)[b]{\smash{\SetFigFont{10}{12.0}{\familydefault}{\mddefault}{\updefault}$M\,[\mbox{GeV}]$}}}
\put(5926,-5911){\makebox(0,0)[lb]{\smash{\SetFigFont{10}{12.0}{\familydefault}{\mddefault}{\updefault}$\mR\!=\!5\,\mbox{GeV}$}}}
\put(6151,-5011){\makebox(0,0)[lb]{\smash{\SetFigFont{10}{12.0}{\familydefault}{\mddefault}{\updefault}$\mR\!=\!1\,\mbox{GeV}$}}}
\end{picture}

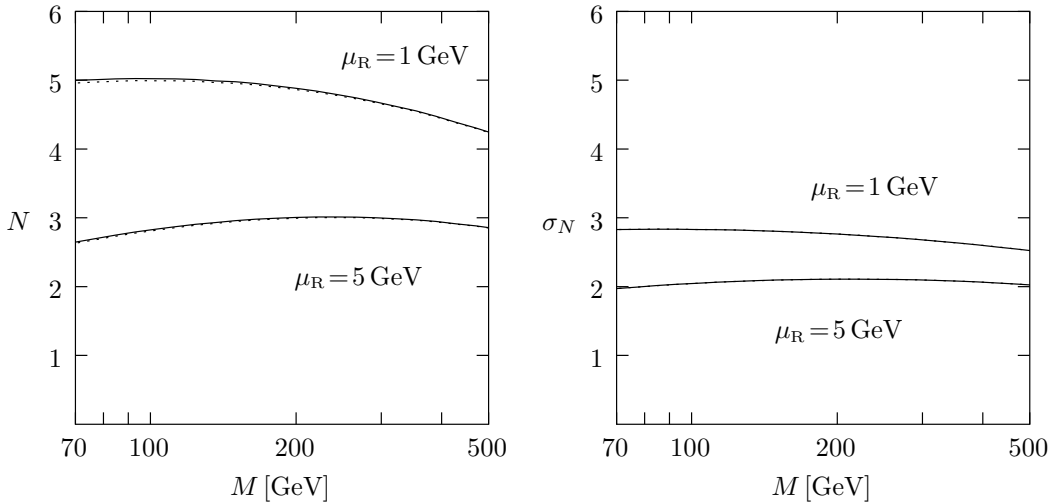
  \caption{The mean value and dispersion of the number of (mini-)jets
    in central Higgs production 
    at LHC for two different resolution scales $\mR$. Solid lines 
    show the SL results up to the 15th order in perturbation theory, 
    dashed lines correspond to the DL approximation.}
\label{fig_higgsjets}}
Here, the DL results give an excellent approximation and 
the SL terms are less significant.
We see that the mini-jet multiplicity and its dispersion are rather
insensitive to the Higgs mass at the energy of the LHC.

\section{Conclusions}\label{sec_conc}
We have derived expressions for the mean and variance of the jet multiplicity
distribution at small $x$, including resummation of leading logarithms of
$x$ and all (leading and sub-leading) logarithms of $Q^2/\mR^2$. 
Our results have been derived using the BFKL equation, but are 
expected to hold in the CCFM formalism as well.

Considered as functions of $\om$, the moment variable Mellin-conjugate
to $x$, our expressions exhibit bad behaviour at large values of $\alb/\om$,
which is associated with the singularity of the leading-order Lipatov
anomalous dimension $\glip$ at $\alb/\om = (4\ln 2)^{-1}$. We would expect this
behaviour to be modified strongly by higher-order corrections.
Although the next-to-leading corrections to $\glip$ are known, a full
calculation of the corresponding corrections to the associated jet
multiplicity has not been performed and would appear much more difficult.

In $x$-space we have only been able to find the jet multiplicity 
and its variance as perturbative expressions. These are obtained 
after expanding the closed expressions in $\om$-space to sufficiently 
high order and applying the inverse Mellin transformation term by term. 
For all practical applications this is sufficient since the perturbative 
series  in $x$-space turns out to be very rapidly convergent.

The multiplicity of mini-jets at small $x$ is an observable that can 
also be computed in Monte Carlo simulations of BFKL dynamics
\cite{MC}. It would be interesting to compare the corresponding 
results with our analytic results. 


\appendix
\section{Lipatov anomalous dimension and characteristic function}
For changing from $\glip$ and its derivatives to $\chi$ and its
derivatives we need the following relations, which
are easily obtained by repeated differentiation of the first:
\begin{eqnarray}
\chi(\glip)  &=&  \frac{\om}{\alb} \\
\chi'(\glip) &=& - \left(\omal\right)^2 \frac{1}{\glip'} \\
\chi''(\glip) &=& \left(\omal\right)^3 \frac{1}{\glip'^3}
          \left( 2 \glip' + \alom \glip'' \right) \\
\chi'''(\glip) &=& \left(\omal\right)^4 \frac{1}{\glip'^5}
          \left[ -6 \glip'^2 - 3 \glip''^2 
                 -6 \alom \glip' \glip''  
                 + \left(\alom\right)^2 \glip' \glip'''  
          \right] \\
\chi''''(\glip) &=& \left(\omal\right)^5 \frac{1}{\glip'^7}
          \left[ 24 \glip'^3 + 36 \alom \glip'^2 \glip''
                + 30 \left(\alom\right)^2 \glip' \glip''^2
                - 8 \left(\alom\right)^2 \glip'^2 \glip'''
          \right.
          \nonumber \\
&& \hspace{2.0cm} \left.
                + 15 \left(\alom\right)^3 \glip''^3 
                - 10 \left(\alom\right)^3 \glip' \glip '' \glip'''
                + \left(\alom\right)^3 \glip'^2 \glip''''
          \right].
\end{eqnarray}
Conversely, the derivatives of $\glip$ are given in terms of
$\chi(\glip)$ by
\begin{eqnarray}
\glip' &=& - \frac{\chi^2}{\chi'} \\
\glip''&=& \frac{\chi^4}{\chi'^3}
           \left( 2 \chi \chi'^2 - \chi'' \right) \\
\glip'''&=& \frac{\chi^6}{\chi'^5}
           \left[ - 6 \chi^2 \chi'^4
                  + 6 \chi \chi'^2 \chi''
                  - 3 \chi''^2 + \chi' \chi''' 
           \right] \\
\glip''''&=& \frac{\chi^8}{\chi'^7}
           \left[ 24 \chi^3 \chi'^6
                  - 15 \chi''^3
                  + 36 \chi \chi'^2 \chi''^2 
                  - 36 \chi^2 \chi'^4 \chi''
                  - 12 \chi \chi'^3 \chi'''
           \right.
           \nonumber \\
&& \hspace{1cm} \left.
                  + 10 \chi' \chi'' \chi'''
                  - \chi'^2 \chi''''
           \right].
\end{eqnarray}

\end{document}